# Tunable Defect Engineering of Mo/TiON Electrode in Angstrom-Laminated HfO$_2$/ZrO$_2$ Ferroelectric Capacitors towards Long Endurance and High Temperature Retention

Sheng-Min Wang, Cheng-Rui Liu, Yu-Ting Chen, Shao-Chen Lee, and Ying-Tsan Tang*

Department of Electrical Engineering, National Central University, Taoyuan 320317, Taiwan, ROC

E-mail: yttang@ee.ncu.edu.tw

**Abstract**

A novel defect control approach based on laminated HfO$_2$/ZrO$_2$ with multifunctional TiN/Mo/TiO$_x$N$_y$ electrode is proposed to significantly improve the endurance and data retention in HZO-based ferroelectric capacitor. The O-rich interface reduces leakage current and prolong the endurance up to $10^{11}$ cycles while retaining a 2Pr value of 34 (μC/cm$^2$) at 3.4MV/cm. Using first-principles calculations and experiments, we demonstrate that the enhancement of endurance is ascribed to the higher migration barrier of oxygen vacancies within the laminated HZO film and higher work function of MoO$_x$/TiO$_x$N$_y$ between top electrode and the insulating oxide. This 2.5-nm-thick TiO$_x$N$_y$ barrier further increase the grain size of HZO, lowering the activation field and thus improving polarization reversal speed. This interfacial layer further decreases the overall capacitance, increases the depolarization field, thereby enhancing the data retention.  By fitting the data using the Arrhenius equation, we demonstrate a 10-year data retention is achieved at 109.6 °C, surpassing traditional SS-HZO of 78.2 °C with a 450°C RTA (required by backend-of-the-line). This work elucidates that interfacial engineering serves as a crucial technology capable of resolving the endurance, storage capability, and high-temperature data retention issues for ferroelectric memory.

## 1. Introduction

Ferroelectric Hf$_{1-x}$Zr$_x$O$_2$ (FE-HZO) has attracted great attention owing to its scalability and programmability for nanoscale devices and has been extensively explored for non-volatile memory applications. In recent years, the combination of HZO with oxide semiconductors, e.g., IGZO and IZO, has been proposed to fabricate metal-ferroelectric-semiconductor field-effect transistors (MFS-FETs). This integration is considered promising for realizing 3D-FeNAND because it exhibits significant switching polarization and high density low power characteristics [1-3]. Compared to FLASH memory with endurance of $10^6$ cycles [4],  FE-HfO$_2$ has been proven to exhibit an endurance exceeding $10^7$ cycles [5, 6]. It is generally believed that the endurance is closely related to defective interface caused by oxidation of electrodes during rapid thermal annealing (RTA) [7, 8]. During switching under an electric field, charge defects may diffuse into the insulator, causing leakage paths [9]. Hence, mitigating defect concentration or suppressing defect diffusion is important for improving the endurances of FE materials [10]. Recent reports show that $O_3$ plasma atomic layer deposition (ALD) can reduce impurities in HZO [11]. This carbon removal lowers the domain pinning effect, improving the wake-up effect of ferroelectricity. Doping $HfO_2$ films with metals having larger atomic radii (e.g., Y, La, Sr) stabilizes ferroelectricity by inducing strain, favoring the orthorhombic phase (O-phase) [12]. A similar concept is applied to the electrode. The thermal expansion coefficients (TECs) of the capping electrode and HZO are different [13]. During the heating and cooling process, the material volume varies, causing in-plane tensile stress [14]. The methods utilize thermal stress to accelerate transformations from the tetragonal (T-phase) to orthorhombic phases (O-phase), thereby enhancing the FE properties. Migita et al. proposed the unbalanced bonding strengths in the nanolaminated film are enhanced at the $HfO_2$ and $ZrO_2$ interfaces [15]. Figure 1(a) and (b) depicts the energy changes of Vo migrations in the solid-solution (SS) and angstrom-laminated (AL)-HZO using first-principles calculations; it is seen that AL-HZO with a periodic stacking of 7.5 Å exhibits a higher energy barrier than SS-HZO, thus suppressing Vo migration [16]. Figure 1(c) displays X-ray photoelectron spectroscopy (XPS) O 1s spectra, the non-latticed oxygen energy peak at 531.9 eV is less dominant in the AL-HZO, revealing AL-HZO exhibits lower Vo content [17]. Kashir et al. proposed W/Pt that accounts for the advantages of both a low TEC and high work function (WF) to extend FE endurance [11]. In this work, we utilize Mo and TiN as the stress layer and diffusion barrier layer (DBL), respectively. Furthermore, Mo, which is susceptible to oxidation (forming $MoO_x$), functions as an oxygen scavenging layer (OSL) to preserve the oxygen concentration near the interface. However, excess O adsorption of the electrode may increase the Vo in HZO, resulting in an increase in leakage current [18]. To address this, we inserted a 2.5-nm TiN electrode between Mo and HZO to hinder excessive oxygen diffusion. Note that the DBL contains O more than 40% (figure S2), suggesting that a transition layer of titanium oxynitride is formed. In this O-rich region, $TiO_xN_y$ acts as a dielectric material [19]. Despite the decrease in 2Pr by the depolarization field $E_{dp} = P_r C_F [\varepsilon_F (C_{IL} + C_F)]^{-1}$ [20] generated by the dielectric $TiO_xN_y$, the superior thermal stability of $TiO_xN_y$ layer [21] significantly enhances the interface, thus improving operation speed and retention time.

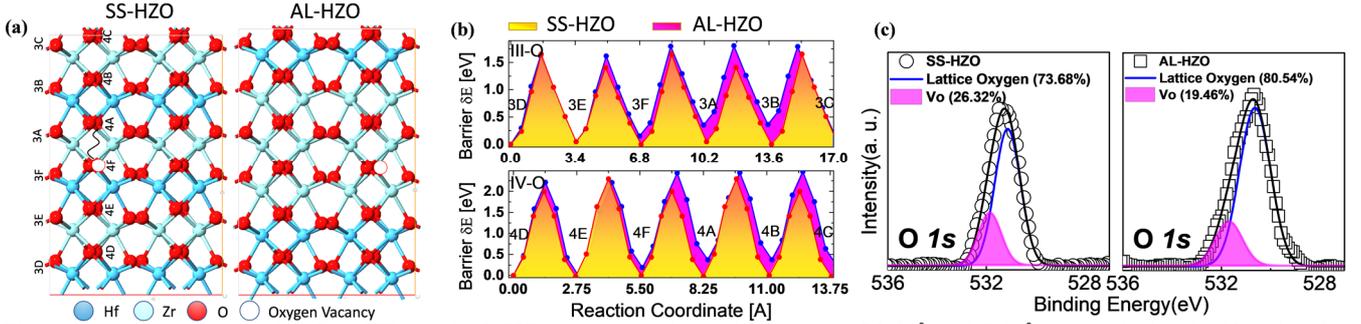

**Figure 1. (a)** Atomic models of SS-HZO and AL-HZO with monolayer thicknesses of 2.5 Å and 7.5 Å, respectively. **(b)** Vo migration barriers along the directions of FE polarization in SS-HZO and AL-HZO, where III-O and IV-O denote the tri-coordinated and tetra-coordinated O sites, respectively. **(c)** XPS O 1s spectra of bulk AL-HZO and bulk SS-HZO.

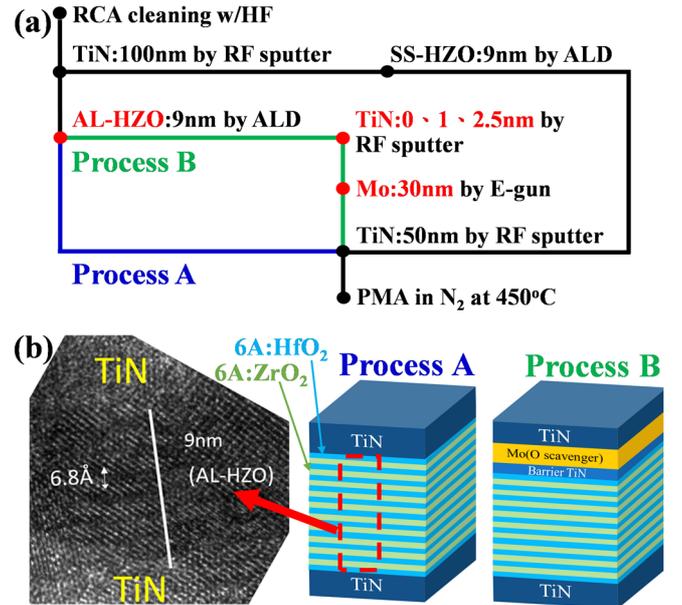

**Figure 2. (a)** Process flows for various FE capacitors. **(b)** Schematic illustrations for process A and B structures, along with cross-sectional TEM image of the AL-HZO stack.

## 2. Experiments

Figure 2 illustrates the fabrication processes of our FE capacitors (FeCAPs). The Si wafer is initially cleaned using the RCA process, followed by deposition of a 100-nm-thick TiN bottom electrode (BE) through RF-sputtering. Subsequently, the $HfO_2$ and $ZrO_2$ layers are cyclically grown by thermal ALD, with each layer being 6.8 Å thick. This results in a total thickness of 9 nm for the HZO. A 50-nm TiN layer is then deposited as the top electrode (TE). Different gate structures, referred to as processes A, B1, B2, and B3, are inserted between the electrodes. In process A, the TE-TiN is directly deposited above and below the HZO layer. In process B, after growing the BE-TiN and HZO layer, a thin TiN layer (B1–B3: 0 nm/1 nm/2.5 nm) is sequentially deposited by RF sputtering. Then, a 30-nm Mo layer is grown using an E-gun, followed by TE-TiN. After patterning completion, both processes A and B undergo post-metallization annealing (PMA) at 450 °C for 30 s. Electrical measurements are then conducted using a Keysight B1500 parameter analyzer. All electric measurements were performed after the wake-up process, by inducing 1000 voltage pulse cycles of 3 V at a frequency of 100 kHz. GI-XRD, HRTEM, and XPS were utilized for the physical structure analyses. First-principles calculations were performed using commercial package (QATK) with GGA-PBE basis. The PseudoDojo exchange potential and nudged elastic band (CI-NEB) were applied to calculate the migration barrier of the Vo. TEC in different electrodes is studied using atomistic simulations based on ab initio molecular dynamics (AIMD), where the FHI-SZP potentials and NPT relaxation were employed to obtain the volumes at different temperatures.

## 3. Results and discussion

*3.1 Laminated Layer, Oxide Scavenger, and TiN Barrier*

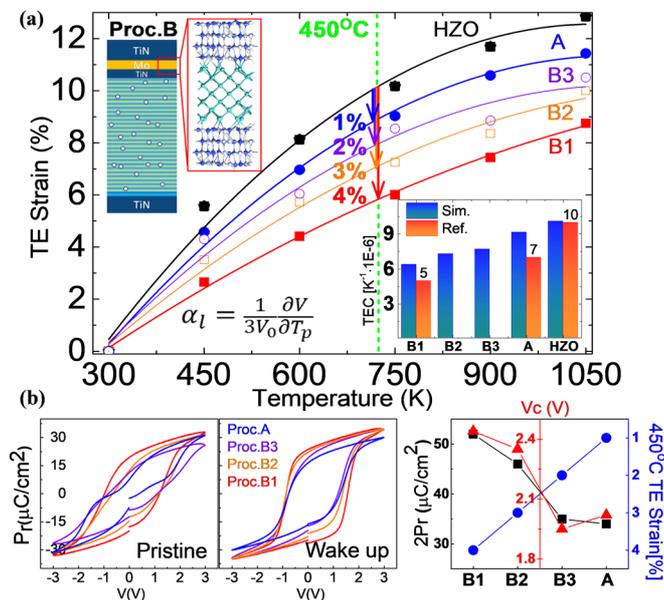

**Figure 3.** (a) Temperature-dependent strains of the HZO and metal gates determined from AIMD; the inset depicts the corresponding TEC. (b) Pristine and wake-up P-V curves for different structures. The change of 2Pr, Vc and 450°C TE strain summarized in right figure

From figure 3, it is observed that process B1 exhibits a highest remnant polarization (Pr) and coercive voltage (Vc) than the other stacking conditions. This is attributed to the stress generated by the TEC different between Mo (~ 4.8 × $10^{-6}$/K) and HZO (~10.1×$10^{-6}$/K) [22]. To validate our experimental results, we conducted first-principles simulations to investigate the temperature-dependent strain for the different gate-stack as shown in figure 3(a). By applying the equation $\alpha_l = \frac{1}{3V_0}\frac{\partial V}{\partial T_p}$ [23], we obtained the TECs of the stackings. The simulation results for TEC indicate that HZO> TiN (9.1×$10^{-6}$/K) > Mo. This trend is in agreement with experimental observations [11]. The in-plane strain is determined by the strain difference at the annealing temperature. In process B1, we observed the largest difference at 450 °C, where the strain reached a maximum of 4%. This stress enhances the proportion of the O-phase within HZO [24][25], leading to maximum Pr and Vc, as depicted in figure 3(b). For increased barrier thickness of TiN (B1-B3), the strain diminishes gradually. Reduced stress lowers T-O transition probability during cooling, yielding more residual T-phase volume, and the wake-up phenomenon appears.

It is reported that a careful trade-off must be considered between the Pr and endurance. Controlling the interface structure crucially governs the ferroelectric performance of the HZO film [26]. Mo capping helps higher Pr while it exhibits poor endurance. The degradation of the endurance might be due to the oxidation of a Mo electrode, which can pull out oxygen from the HZO layer and increase the defect concentration in insulating HZO film. Figure 4(a) presents the Hf 4f spectrum. The sub-oxide energy peaks at 17.57 eV and 19.23 eV are more prominent, revealing that direct Mo capping exhibits a higher defect concentration than the TiN electrode [27]. Although the WF of $MoO_x$ (~6.3 eV) [28] (Supporting information) is higher than that of TiN (~4.7 eV), the charged Vo may introduce mid-gap states in the HZO, increasing the leakage current, as illustrated in figure 4(b) [29]. To mitigate this, we inserted a 2.5-nm TiN between Mo and HZO, which effectively reduce excess oxidation in Mo and Vo formation within HZO, as shown in figure 4c (process B3). With increasing TiN barrier thickness, the high WF of $MoOx(6.3)/TiO_xN_y(5.4)$ [30] is less dominant, and the work function is reduced to $Mo(4.7)/TiN(4.7)/TiO_xN_y(5.4)$. This leads to a decrease in the Schottky barrier and a subsequent increase in leakage current. Thus, it is important to control the thickness of barrier TiN. The O mapping in figure 4(c) and XPS depth analysis in figure S2 demonstrate that the 2.5-nm TiN experiences significant oxidation to form $TiO_xN_y$ in the presence of the OSL. Next, to understand the impact of barrier $TiO_xN_y$ on fatigue, we investigated endurance under various stacking configurations. Figure 4(d) illustrates the stress tests using a trapezoidal waveform of voltage ±3 V and pulse width 1 μs. Obviously, AL-HZO exhibits superior endurance than SS-HZO by three orders of magnitude. This is due to the higher Vo migration barrier in AL-HZO than SS-HZO [16]. In addition, the B3 gate-stack outperforms the A gate-stack in endurance by two orders of magnitude, enduring up to $10^{11}$ cycles. This improvement is attributable to two factors. Firstly, an O-rich interfacial layer of $TiO_xN_y$ ensures stoichiometry of HZO. Secondly, the oxidation of 2.5 nm-TiN and Mo possesses a higher WF ($MoO_x$/$TiO_xN_y$), thus inhibits charge tunnelling across the Schottky barrier to ensure a lower chance of breakdown, thereby enhancing the endurance.

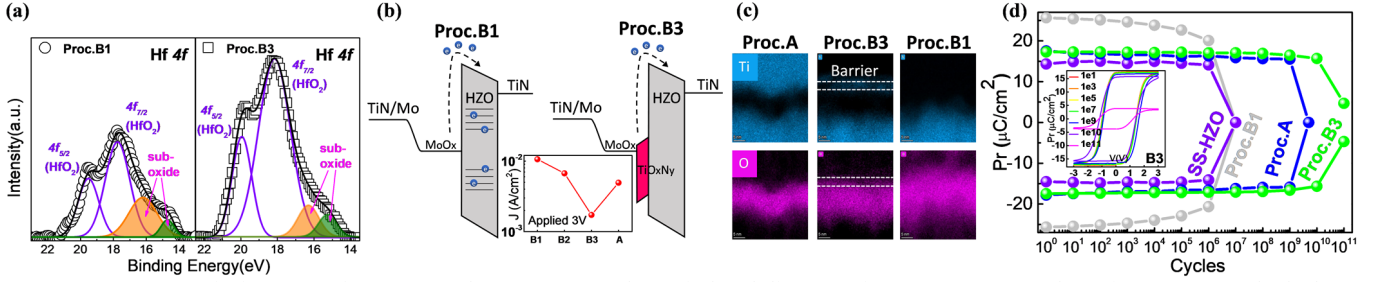

**Figure 4. (a)** XPS Hf 4f spectra of Proc.B1 and Proc.B3 **(b)** Schematic band diagrams for processes B1 and B3; inset presents the leakage current at 3 V. **(c)** EDX mapping profiles for different gate-stack structures, with the dashed line representing the TiN barrier. **(d)** Endurances of different gate-stack configurations, with the inset displaying the cycling-number-dependent P-V curves for process B3.

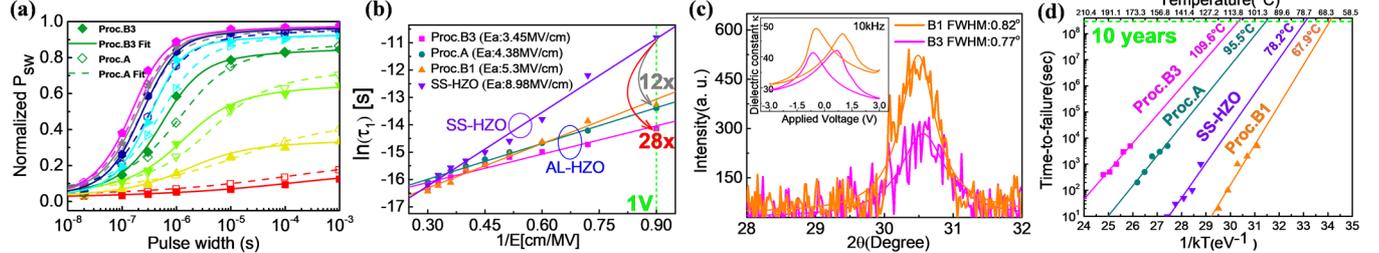

**Figure 5. (a)** Switching kinetics of processes A and B3. **(b)** Switching time variation with the applied external voltage. **(c)** GI-XRD data of processes B1 and B3; the inset shows the total capacitance κ-V curves at 10 kHz. **(d)** High-temperature accelerated method is employed to perform data retention tests on various FeCAP devices.

### 3.2 *Domain Switching Kinetics Under Different Electrodes*

Figure 5(a) depicts the polarization switching kinetics for different voltages and pulse widths. The methodology employed here involves prepolling the dipoles to the same direction using a 3 V, 1 μs pulse. Subsequently, the program voltages (0.25–3 V) and pulse widths ($2\times10^{-8}$ to $10^{-3}$ s) are applied. The polarizations are then read using the PUND measurements. Next, we adopted the nucleation-limited switching (NLS) model proposed by Jo et al. to extract the switching time [31].

$$\Delta P(t) = 2P_s \int_{\infty}^{\infty}\left[1 - exp\left\{-\left(\frac{t}{t_0}\right)^n\right\}\right]\left\{\frac{A\omega/\pi}{(\log t_0 - \log \tau_1)^2 + \omega^2}\right\} dlog t_0$$

Figure 5(b) depicts the switching time versus electric field trend. According to Merz's law, $\ln(\tau_1) = \ln(\tau_1') + E_a/E$ [32], where the slope denotes the activation field ($E_a$). We fitted the experimental data and observed $E_a^{SS:HZO} > \bar{E}_a^{B1} > \bar{E}_a^{A} > \bar{E}_a^{B3}$, where the bar indicates AL-HZO. Remarkably, the activation field of AL-HZO is nearly half that of SS-HZO and a significant (12 times) enhancement in switching speed was found at 1 V. Furthermore, after incorporating the 2.5-nm TiN DBL, it can be seen that Ea is additionally reduced 20%, from 5.3 MV/cm to 3.45 MV/cm, leading to 28 times overall increase in the polarization reversal speed. It is noteworthy that our data for Ea in SS-HZO is similar to the previous report [33]. The accelerated behavior can be attributed to both the grain size and the depolarization field. We first analyzed the grain size in AL-HZO films using GIXRD. From figure 5(c) we scan the O/T characteristic peak near 2θ=30.4°. According to Ref [34], the average grain size radius can be defined by $r = \frac{0.9\lambda}{W\cos\theta}$, where λ is the X-ray wavelength and W is the full-width at half maximum (FWHM). Since $W_{B3} < W_{B1}$, B3 exhibits a larger orthorhombic grains compared to B1. Nonetheless, due to the slight rightward shift of the highest point, B3 exhibits characteristics slightly closer to an anti-ferroelectric T-phase, which agrees with the initial value of 2Pr (~34μC/$cm^2$) observed in Fig.3(b). Additionally, due to $E_c \propto r^{-2/3}$ [35-37], the larger the grain size, the smaller the Ec value. This result is also consistent with P-V curves in figure 3(b). Larger grain has a faster polarization switching operation [38]. According to Landau-Ginsburg-Devonshire (LGD) theory for 1st order FE and Jo [28], $E_a = \frac{UE_0}{k_BT}$, where $U \propto (\alpha + 2\gamma)^2/4\beta$. Here, α, β, and γ are the parameters of the Landau free-energy equation [39-41]. For FE materials, α<0 and γ>0, γ represent the depolarization factors from interfacial layers. Despite the higher T-phase (a larger dielectric constant compared to O-phase) in process B3, the κ-V curves shown in Fig. 5(c) contradict this observation. This discrepancy may arise from not considering the presence of $C_{IL}$. In other words, oxidation of the DBL ($TiO_xN_y$) decreases the $C_{IL}$, resulting in a lower overall capacitance. The reduced $C_{IL}$ then increases $E_{dp}$ (ratio: γ), to obtain a smaller $U$, thereby enhancing the switching speed.

## 3.3 *Retention Time Under Different Electrodes*

Figure 5(d) presents the retention properties at different temperatures for different capacitors. By fitting the data using the Arrhenius equation, we extrapolated the maximum temperature for 10 years of data retention [42]. Among these structures, process B3 demonstrates superior thermal stability, with a retention temperature near 109.6 °C, surpassing SS-HZO (78.2 °C). Note that a BEOL-compatible annealing temperatures of 450 °C is applied here for crystallization of the HZO thin film. This enhanced stability may stem from the low thermal resistance of superlattice (SL) with coherent phonon transport [43-45]. Nevertheless, the oxidation of the high-temperature Mo TE layer may cause defects in HZO, consequently reducing the reliability of the insulator. From this perspective, the B3 configuration will have much more promising in the ferroelectric memory.

## 3.4 *Reliable Triple-level cell Operation*

The accumulation of charged defects and charge trapping/de-trapping processes at the interface between the electrode and HZO are the primary root causes of polarization fatigue and Pr variation after field cycling [46]. Reducing the overlap between each intermediate polarization state is crucial for multi-level cell (MLC) operation [47-49]. Figure 6(a) illustrates the endurance data of process A and process B3 under eight different write biases (from -0.5 to -2 V). The polarization is read after every two orders of magnitude. All states in B3 exhibit stable endurances up to $10^9$ cycles, while states in A show significant Pr degradation after $10^7$ cycles.

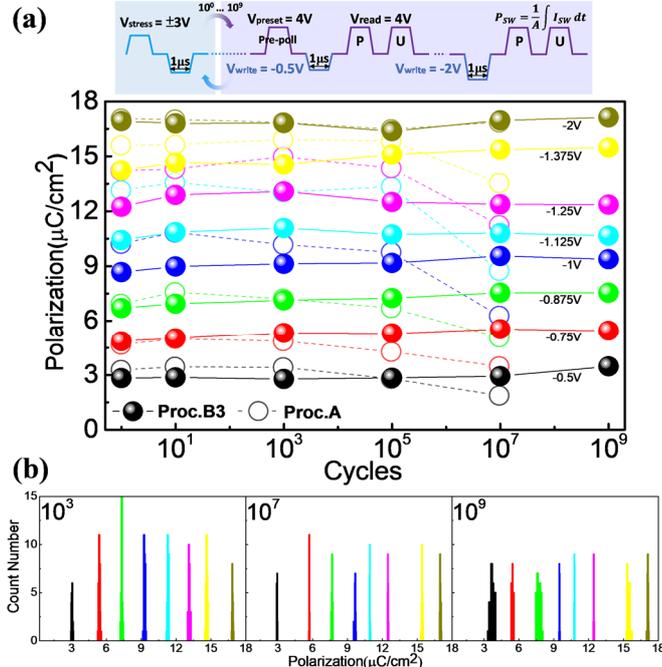

**Figure 6.** (a) Cycling operation for TLC using a unipolar pulse scheme. (b) CtC variation of polarization for each state in process B3 after $10^3$, $10^7$ and $10^9$ cycles with 30 repetition

*Table 1: Benchmarking of various MLC technologies.*

|  | Ref [42] | Ref [43] | Ref [44] | Ref [45] | This work |
|---|---|---|---|---|---|
| **TE** | W | TiN | W | Pt/TiN | **TiN/Mo/2.5nm-TiN** |
| **FE Material** | SS-HZO/ZrO₂/SS-HZO | AlON/SS-HZO/HfO₂ | SL-HZO | Si:HfO₂ | **AL-HZO** |
| **BE** | W | TiN | W | TiN | **TiN** |
| **RTA** | 600 ºC | 400 ºC | 600 ºC | 600 ºC | **450 ºC** |
| **Max 2Pr** | ~34 µC/cm² | ~23 µC/cm² | ~20 µC/cm² | ~30 µC/cm² | **~34 µC/cm²** |
| **Write Pulse Width/ Amp.** | 10µs/ -3V~ +3V | 1µs/ -4V~ +4V | 4µs/ -2V~ +2V | 2µs/ +0.9V~ +3V | **1µs/ -0.5V~ -2V** |
| **Storage Capability (states)** | MLC (4) | TLC (8) ~10⁸ | MLC (4) | TLC (8) | **TLC (8) ~10⁹** |
| **Endurance** | ~10⁷ | ~10⁸ | ~10⁶ | N/A | **~10¹¹** |
| **Decade Retention Temp.** | N/A | N/A | ~85ºC | ~85ºC | **~109 ºC** |

Figure 6(b) demonstrates the cycle-to-cycle (CtC) variation of Pr for the eight states by repeating each voltage pulse 30 times. It can be seen that the eight states can be well distinguished without read disturbance even at $10^9$ cycles. This reflects the presence of inevitable defects in traditional TiN electrode on HZO (process A), while a relatively high Schottky barrier in process B3 suppresses the charge injection processes, thereby improving endurance. Table 1 shows benchmark results for various MLC FeCAPs. The sandwiched TiN/Mo/2.5nm-TiN gate-stack shows competitive advantages in BEOL compatible process, good remanent polarization value, low operation voltage, faster operation speed, higher storage capability, more robust endurance cycle for each state and high temperature data retention [50-53].

## 4. Conclusion

We conducted first-principles calculations and experiments to systematically study the influences of OSL, TiN barrier, and AL-HZO. AL-HZO effectively reduces the Vo migration barrier due to an unbalanced bond between HfO₂ and ZrO₂, thus improving endurance by three orders of magnitude compared to traditional solid solution HZO. The OSL/TiN barrier with a high work function ($MoO_x/TiO_xN_y$) minimizes charge trapping, ensuring a reliable interface close to the TE, and resulting in a read/write speed improvement of 28 times and an endurance exceeding $10^{11}$ cycles. It is demonstrated that MFM capacitors with TLC storage can be achieved with discernible eight states for up to $10^9$ cycles and a decade retention at temperatures exceeding 109 ºC. This paves the way for enabling high thermal stability memory applications. Therefore, adopting OSL/TiN top electrode is highly promising gate-stack in semiconductor memory.

## Acknowledgements

This work was supported by the National Science and Technology Council (NSTC) of Taiwan (No. 111-2218-E-A49-016-MBK, 112-2622-8-A49-013 -SB, 109-2221-E-008-093-MY3,112-2119-M-A49-006) and Taiwan Semiconductor Research Institute (No. JDP112-Y1-023).